\documentclass[12pt]{article}
\usepackage{amsfonts}
\usepackage{geometry}
\usepackage[doublespacing]{setspace}

\input{tcilatex}
\begin{document}

\begin{center}
{\huge A proof of Bell's inequality in quantum mechanics using causal
interactions}

\bigskip 

James M. Robins, Tyler J.\ VanderWeele

Departments of Epidemiology and Biostatistics, Harvard School of Public
Health

\bigskip

Richard D. Gill

Mathematical Institute, Leiden University

\bigskip 

\noindent \textbf{Abstract}
\end{center}

We give a simple proof of Bell's inequality in quantum mechanics which, in
conjunction with experiments, demonstrates that the local hidden variables
assumption is false. The proof sheds light on relationships between the
notion of causal interaction and interference between particles.

\bigskip


\bigskip \pagebreak

\noindent \textbf{Introduction}

\bigskip

Neyman introduced a formal mathematical theory of counterfactual causation
that now has become standard language in many quantitative disciplines,
including statistics, epidemiology, philosophy, economics, sociology, and
artificial intelligence, but not in physics. Several researchers in these
disciplines (Frangakis et al., 2007; Pearl, 2009) have speculated that there
exists a relationship between this counterfactual theory and quantum
mechanics, but have not provided any substantive formal relation between the
two. In this note, we show that theory concerning causal interaction,
grounded in notions of counterfactuals, can be used to give a
straightforward proof of a result in quantum physics, namely, Bell's
inequality. Our proof relies on recognizing that results on causal
interaction (VanderWeele, 2010) can be used to empirically test for
interference between units (VanderWeele et al., 2011). It should be stressed
that a number of extremely short and elegant proofs of both Bell's original
inequality (and its generalizations) are already available in the physics
literature. In fact some of these proofs are based on reasoning with
counterfactuals (Gill et al., 2001). Our contribution is to explicitly show
relations to the theory of causal interactions.

We motivate our proof with an exceedingly short history of the Bell
Inequality that is elaborated upon later. A non-intuitive implication of
quantum theory is that pairs of spin 1/2 particles (e.g., electrons) can be
prepared in an entangled state with the following property. When the spins
of both particles are measured along a common (spatial) axis, the
measurement of one particle's spin perfectly predicts the spin of the other;
if the first particle's spin is up, then the spin of the second must be
down. One explanation would be that the measurement itself of the first
particle determined the spin of the second, even if physically separated,
perhaps, by many light years. This would mean that reality was not "local";
what occurred at one place would affect reality (i.e. the spin of the second
electron) at another. However, Einstein believed in "local realism" and
argued that the more plausible explanation was that both particles are
carrying with them from their common source 'hidden' correlated spin
outcomes which they will exhibit when measured (Einstein et al., 1935). \ He
therefore argued for "local realism" and rejected the previous explanation.
Bohr disagreed with Einstein and his "local realist" assumption. Neither
Einstein nor Bohr apparently realized that the hypothesis of local realism
was subject to empirical test. In 1964, John Bell showed that a test was
possible; he proved that if strict locality were true, there would be
certain inequality relations between measurable quantities that must hold
(Bell, 1964); quantum theory predicted that these inequalities must be
violated. Experiments found Bell's inequalities were indeed violated (though
see discussion below for further comments). Einstein was wrong; local
realism is false.

\bigskip

\noindent \textbf{A Proof of Bell's Inequality Using Causal Interactions}

\bigskip

We now show how results on causal interaction can be used to produce an
alternative proof of Bell's theorem. Suppose we have two particles and can
use devices to measure the spin of each, along any axis of our choosing. Let 
$X_{1}$ and $X_{2}$ be two "interventions" each taking values in $\{0,1,2\}$%
, where $X_{1}$ records the angle (i.e. axis in space) at which particle 1
is measured, and $X_{2}$ records the angle at which particle 2 is measured,
and where $0,1,2$ correspond to three particular angles. Let $%
Y_{1}(x_{1},x_{2})$ be the binary spin (up$=1$ or down$=-1$) of particle 1
and $Y_{2}(x_{1},x_{2})$ be the spin for particle 2, when particle 1 is
measured at angle $x_{1}$ and particle two is measured at $x_{2}$. In the
language of the Neyman model $Y_{i}(x_{1},x_{2})$ is the counterfactual
response of particle $i$ under the joint intervention $(x_{1},x_{2})$. Let $%
M(x_{1},x_{2})=1\{Y_{1}(x_{1},x_{2})=Y_{2}(x_{1},x_{2})\}$ be an indicator
function that the spin directions agree so that $M(x_{1},x_{2})=1$ if the
spin direction agree and $M(x_{1},x_{2})=0$ if they disagree. Suppose that
the particles are in a maximally entangled state. Then, according to quantum
mechanics of the 2 particle system, for $i,j=0,1,2,$ $E[M(x_{1}=i,x_{2}=j)]=%
\sin ^{2}(\Delta _{ij}/2)$, where $\Delta _{ij}$ is the angle between angles 
$i$ and $j$. This result has been confirmed by experiments in which the
angles of measurement were randomized (though see discussion below for
further comments). Therefore in what follows we take $%
\{E[M(x_{1},x_{2})];x_{1}\in \{0,1,2\},x_{2}\in \{0,1,2\}\}$ as known, based
on the data from experiment. Since $\sin \left( 0\right) =0$, $%
M(i,i)=0,i=0,1,2$, with probability $1$ and, therefore, also $%
Y_{1}(i,i)=-Y_{2}(i,i)=0$, with probability $1$, as mentioned earlier.

We formalize the hypothesis of "local hidden variables" by the hypothesis
that spin measured on one particle does not depend on the angle at which the
other particle is measured. This can be stated as: for all angles $%
(x_{1},x_{2})$

\qquad $Y_{1}(x_{1},x_{2})=Y_{1}(x_{1})$

\qquad $Y_{2}(x_{1},x_{2})=Y_{2}(x_{2})$.

In some of the experiments referenced above the times of the two
measurements were sufficiently close and the separation of the particles
sufficiently great that even a signal traveling at the speed of light could
not inform one particle of the result of the other's spin measurement.
Therefore, refuting the hypothesis of "local hidden variables" implies
reality is not local and therefore we can essentially treat the hypothesis
of local hidden variable and local reality as the same; we return to this
point in the discussion.

The hypothesis asserts both locality and reality. It asserts locality
because the angle $x_{2}$ at which particle $2$ is measured has no effect
the spin $Y_{1}(x_{1})$ of particle 1$.$ It asserts reality because the spin 
$Y_{i}(x)$ of a particle measured along axis x is assumed to exist for every 
$x,$even though for each $i$, only one of the $Y_{i}(x)$ is observed; the
one corresponding to the axis along which particle was actually measured.
All other $Y_{i}(x)$ are missing data in the language of statisticians or,
equivalently, hidden variables in the language of physicists. The
counterfactuals $Y_{i}(x)$ correspond exactly to what Einstein called
"elements of reality". In the language of counterfactual theory, the
hypothesis of local reality is, by definition, the hypothesis of no
interference between treatments. In the following a unit may be taken to be
a pair of entangled particles.

\bigskip

\textbf{Theorem 1}. If for some unit, $M\left( 0,0\right) =0$, $M(1,2)=1$, $%
M(0,2)=0$, $M(1,0)=0$ then the hypothesis of 'local hidden variables' is
false.

\bigskip

Proof. By contradiction: Suppose the hypothesis holds. Now $M(1,2)=1$
implies either (a) that $Y_{1}(1)=Y_{2}(2)=1$ or (b) that $%
Y_{1}(1)=Y_{2}(2)=-1$. Suppose that (a) holds: then $M(0,2)=0$ and $%
Y_{2}(2)=1$ imply $Y_{1}(0)=-1$. But, $M(1,0)=0$ and $Y_{1}(1)=1$ imply $%
Y_{2}(0)=-1$ and thus,by $M\left( 0,0\right) =0$, that $Y_{1}(0)=1$, a
contradiction.

\bigskip

Suppose instead that (b) holds. Then $M(0,2)=0$ and $Y_{2}(2)=-1$ implies $%
Y_{1}(0)=1$. But $M(1,0)=0$ and $Y_{1}(1)=-1$ implies $Y_{2}(0)=1$ and thus,
by $M\left( 0,0\right) =0$, that $Y_{1}(0)=-1$, again a contradiction. Thus
it cannot be the case that $Y_{i}(x_{1},x_{2})=Y_{i}(x_{i}),i=1,2.$

The next result is given in VanderWeele (2010) in the context of testing for
a causal interaction, sometimes referred to as "epistasis" in genetics. It
relates the empirical data $E[M(x_{1},x_{2})]$ to the existence of a unit
satisfying $M(1,2)=1,M(0,2)=M(1,0)=M(0,0)=0$. Within the counterfactual
framework, this would constitute a causal interaction for the variable $M$.
Since the proof of the result relating observed data $E[M(x_{1},x_{2})]$ to
units such that $M(1,2)=1,M(0,2)=M(1,0)=M(0,0)=0$ is essentially one line,
we give it here also for completeness.

\bigskip

\textbf{Theorem 2}. If $E[M(1,2)]-E[M(0,2)]-E[M(1,0)]-E[M(0,0)]>0$, then
there must exist a unit with $M(1,2)=1,M(0,2)=M(1,0)=M(0,0)=0$.

\bigskip

Proof. By contradiction. Suppose there were no unit with $%
M(1,2)=1,M(0,2)=M(1,0)=M(0,0)=0$. Then, for all units, $%
M(1,2)-M(0,2)-M(1,0)-M(0,0)\leq 0$ which implies $%
E[M(1,2)]-E[M(0,2)]-E[M(1,0)]-E[M(0,0)]\leq 0$, a contradiction.

\bigskip

An immediate corollary of Theorems 1 and 2 is then:

\bigskip

\textbf{Corollary}. If $E[M(1,2)]-E[M(0,2)]-E[M(1,0)]-E[M(0,0)]>0$, then the
the hypothesis of 'local hidden variables' is false.

\bigskip

This corollary is referred to as Bell's theorem in the physics literature.
Its premise is referred to as Bell's inequality. As noted above, from the
quantum mechanics of the 2-particle system, and confirmed by experiment, $%
E[M(x_{1}=i,x_{1}=j)]=\sin ^{2}(\Delta _{ij}/2)$. Thus we have that:

\[
E[M(1,2)]-E[M(0,2)]-E[M(1,0)]-E[M(0,0)]=\sin ^{2}(\Delta _{12}/2)-\sin
^{2}(\Delta _{02}/2)-\sin ^{2}(\Delta _{10}/2)-0 
\]%
From this it follows that the local hidden variables assumption is rejected
if%
\[
\sin ^{2}(\Delta _{12}/2)>\sin ^{2}(\Delta _{02}/2)+\sin ^{2}(\Delta
_{10}/2) 
\]%
but the angles $0,1,2$ can easily be chosen to satisfy this inequality. Thus
the hypothesis of local hidden variables\ is false.

The prototypical Bell inequality, and accompanying experiment, has in recent
years spawned a multitude of variations involving more than two particles,
measurements with more than two outcomes, and more than two possible
measurements at each location; see for instance Zohren, et al. (2010) for a
striking version of "Bell" obtained simply by letting the number of outcomes
be arbitrarily large. Popular inequalities and experiments are compared in
terms of statistical efficiency by Van Dam et al. (2005). Other connections
to statistics (missing data theory) and open problems are surveyed in Gill
(2007).

\bigskip

\noindent \textbf{Discussion}

\bigskip

We claimed above that there were experimental results that violated Bell's
inequality and therefore ruled out local hidden variables. However, there
remains several small possible loopholes. Perhaps the most important one of
which is the following: in these experiments for every entangled pair that
we measure we often fail to detect one of the two particles because the
current experimental set-up is imperfect. The experimental results we noted
above are actually the results conditional on both particles' spins being
measured. If those pairs were not representative of all pairs, that is, if
the missing pairs are not missing at random, it is logically possible that
the experimental results can be explained by local hidden variables where
the values of $Y_{1}(x_{1})$ and $Y_{2}(x_{2})$ also determine the
probability of the spin of both being observed. The results of experiments
that close this loophole by observing a higher fraction of the pairs should
be available within the next several years. Nearly all physicists believe
that the results of these experiments will be precisely as predicted by
quantum mechanics and thus violate Bell's inequality.

Henceforth, we assume Bell's inequality is violated and that we have
therefore ruled out local hidden variables. We now return to the question of
whether this rules out local reality. As noted above, experiments have been
conducted such that the times of the two measurements were sufficiently
close and the separation of the particles sufficiently great that even a
signal traveling at the speed of light could not inform one particle of the
result of the other's spin measurement. Since physical signals cannot be
transmitted faster than the speed of light, the effect of the measurement of
the first particle on the outcome of the second cannot be explained by any
physical mechanism. Therefore ruling out local hidden variable would also
effectively rule out local reality.

Since the hypothesis of local reality is false, we conclude that the
alternative is true and angle at which particle 1 is measured has a causal
effect on the spin of particle 2. Note, even under the alternative, we have
assumed that $Y_{1}(x_{1},x_{2})$ exists for all $(x_{1},x_{2})$. Thus our
assumption of 'reality' remains; the hypothesis that "reality" is local has
been rejected. However, quantum mechanics is generally assumed to be
irreducibly stochastic. We could have accommodated this assumption by
positing stochastic counterfactuals $p_{1}(x_{1},x_{2})$ and $%
p_{2}(x_{1},x_{2})$ defined for all $(x_{1},x_{2})$ with the measured spin $%
Y_{i}(x_{1},x_{2})$ being the realization of a Bernoulli random with success
probability $p_{i}(x_{1},x_{2})$. That is, we could assume that the elements
of reality are the counterfactual probabilities $p_{i}(x_{1},x_{2})$. Our
hypothesis of stochastic locality is then $p_{1}(x_{1},x_{2})=p_{1}(x_{1})$
and $p_{2}(x_{1},x_{2})=p_{2}(x_{2})$. The proof given above, again combined
with the experimental results, can be used to reject this hypothesis by
using a coupling argument as in VanderWeele and Robins (2011).

A perhaps more radical point of view is that is often attributed to the
Copenhagen school: the mathematical theory of quantum mechanics successfully
predicts the results of experiments, without positing any "elements of
reality" (counterfactuals), even the above non-local stochastic ones. Thus
the question of their existence is not a scientific question, as it is not
subject to empirical test and our most successful scientific theory, quantum
mechanics, has no need of them. This is appealing to physicists because it
restores locality in the following sense. To become entangled two particles
must interact and this interaction, even in the laws of quantum mechanics,
occurs locally. Entanglement leads to correlated measurements. Once
entangled, these correlations will persist irrespective of the particles'
separation as described earlier. However, following the Copenhagen school,
to say counterfactuals $Y_{i}(x_{1},x_{2})$ do not exist is to say that
question of whether the measurement of the spin of particle 1 had an effect
on the spin on particle 2 cannot be asked; not every event has a cause. In
all physical theories prior to quantum theory, it was possible to imagine,
alongside the actual measurements of actual experiments, what would have
been observed had we done something differently (i.e. counterfactuals),
while still preserving locality. This is not possible with quantum
mechanics. In summary, Bell's inequality (and its experimental support) show
that the Copenhagen standpoint of abandoning counterfactuals is not only
possible, it is also necessary to take this standpoint if we want to retain
"locality" as a fundamental part of our world picture.

\bigskip 

\noindent \textbf{Acknowledgements.} This research was funded by NIH\ grant
ES017876.

\bigskip

\noindent \textbf{References}

\bigskip

\noindent Bell, J.S. (1964). On the Einstein Podolsky Rosen paradox. Physics
1:195-200.

\bigskip

\noindent Einstein, A., Podolsky, B. and Rosen, N. (1935). Can
quantum-mechanical description of physical reality be considered complete?
Physical Review, 47:777-780.

\bigskip

\noindent Frangakis, C.E., Rubin, D.B., An, M.W. and MacKenzie, E. (2007).
Principal stratification designs to estimate input data missing due to death
(with discussion). Biometrics, 63:641-662.

\bigskip

\noindent Gill, R.D. (2007).Better Bell inequalities (passion at a
distance). IMS Lecture Notes Monograph Series, Vol. 55, 135-148.

\bigskip

\noindent Gill, R.D., Weihs, G., Zeilinger, A., and Zukowski, M. (2001). No
time loophole in Bell's theorem: The Hess-Philipp model is nonlocal.
Proceeding of the\ National Academy of Sciences, 99:14632-14635.

\bigskip

\noindent Pearl, J. (2009). Causality: Models, Reasoning, and Inference.
Cambridge: Cambridge University Press, 2nd edition.

\bigskip

\noindent van Dam, W., Grunwald, P. and Gill, R.D. (2005). The statistical
strength of nonlocality proofs. IEEE-Transactions on Information Theory 51,
2812-2835.

\bigskip

\noindent VanderWeele, T.J. (2010). Epistatic interactions. Statistical
Applications in Genetics and Molecular Biology, 9, Article 1:1-22.

\bigskip

\noindent VanderWeele, T.J. and Robins, J.M. (2011). Stochastic
counterfactuals and sufficient causes. Statistica Sinica, in press.

\bigskip

\noindent VanderWeele, T.J., Tchetgen Tchetgen, E.J., and Robins, J.M.
(2011). A mapping between interactions and interference: implications for
vaccine trials. Technical Report.

\bigskip

\noindent Zohren, S., Reska, P., Gill, R.D., and Westra, W. (2010). A tight
Tsirelson inequality for infinitely many outcomes. Europhysics Letters
90:10002.

\end{document}